# Formation of $Mn^{2+}$ in $La_{2/3}Ca_{1/3}MnO_3$ Thin Films due to Air Exposure

S. Valencia*, A. Gaupp and W. Gudat
BESSY, Albert-Einstein-Str. 15, D-12489, Berlin, Germany

Ll. Abad, Ll. Balcells, A. Cavallaro and B. Martínez
Institut de Ciència de Materials de Barcelona, Campus de la UAB, E- 08193 Bellaterra, Spain

F. J. Palomares
Instituto de Ciencia de Materiales de Madrid, CSIC, Cantoblanco, E-28049 Madrid, Spain

**Abstract**

We observe a Mn valence instability on $La_{2/3}Ca_{1/3}MnO_3$ thin films, grown on top of $LaAlO_3$ (001) substrates, by X-ray absorption spectroscopy at the Mn L-edge and O K-edge. As grown samples, in-situ annealed at 800°C in oxygen, exhibit a Curie temperature well below that of the bulk material. Upon air exposure they develop a $Mn^{2+}$ spectral signature, in addition to the expected $Mn^{3+}$ and $Mn^{4+}$, contributions which increases with time. Simultaneously a further reduction of the saturation magnetization, $M_S$, of the films is detected. The similarity of the spectral results obtained by total electron yield and fluorescence yield spectroscopy indicates that the location of the Mn valence anomalies is not confined to a narrow surface region of the film, but can extend throughout the whole thickness of the sample. High temperature annealing at 1000°C in air, immediately after growth, improves the magnetic and transport properties of such films forwards the bulk values and the $Mn^{2+}$ signature in the spectra disappears. The Mn valence is then stable even to prolonged air exposure. We propose a mechanism for the $Mn^{2+}$ ions formation and discuss the importance of these observations with respect to previous findings and production of thin films devices.

PACS number(s): 78.70.Dm, 75.47.Lx, 75.70.-i, 71.20.-b

*Corresponding author, e-mail: Valencia@bessy.de

# I. Introduction

Rare-Earth Manganese perovskites have extensively been studied to understand the origin of the observed colossal magnetoresistance (CMR) properties, i.e. the strong correlation between their magnetic and transport properties, because of their enormous potential for applications. As outcome of these investigations, a complex scenario has been suggested where in addition to the double exchange theory [1-3] and the Jahn Teller distortions of the lattice [4], charge and orbital degrees of freedom as well as a natural tendency of the material towards phase separation is taken into account [5-7] in order to describe the observed behaviour.

The implementation of magnetoelectronic devices based on these mixed valence oxide materials requires in most of the cases the use of thin films. Even though the preparation techniques of oxide thin films have experienced a tremendous advance recently, there are still serious problems regarding magnetotransport properties, since bulk-like behaviour is only achieved under specific conditions. Structural mismatch with the substrates [8], inhomogeneities located at interfaces [9], surface segregation [10-12] and oxygen depletion [13], are only some of the possible explanations offered so far. Until now, not much attention has been paid to the possible role of the long-term valence stability of Manganese ions in thin films when affected by external environmental conditions. Brousard et al. [14] reported on the stability of the manganese valence on the time scale of some minutes as a result of air exposure for $La_{2/3}Ca_{1/3}MnO_3$ (LCMO) on $SrTiO_3$ (STO) substrates. Hundley et al. [15] suggested the presence of $Mn^{2+}$ in polycrystalline $La_{1-x}Ca_xMnO_{3+\delta}$ samples in order to explain the disagreement between their temperature-dependent thermoelectric power results and what would be expected from the nominal $Mn^{3+}/Mn^{4+}$ composition. More recently, by means of X-ray absorption spectroscopy, de Jong et al. [16] found first experimental evidence of the existence of divalent Mn at the surface of $La_{2/3}Sr_{1/3}MnO_3$ (LSMO) thin films on STO substrate.

In previous X-ray absorption spectroscopy (XAS) experiments on air-exposed LCMO thin films grown on LAO substrates, we observed the presence of $Mn^{2+}$ ions, its spectral contribution increasing with time. We considered three possible explanations, i) an oxygen removal from the sample due to the ultra high vacuum (UHV) ($10^{-8}$ mbarr) of the experimental chamber, ii) a damaging of the films because of the synchrotron radiation and iii) an ambient atmosphere induced $Mn^{2+}$ formation. In order to answer the question about the $Mn^{2+}$ origin and its effect on the magnetic properties of the films we performed a systematic investigation by means of X-ray absorption spectroscopy of the long-term stability of the Mn valence in a set of LCMO thin films grown on LaAlO$_3$ (LAO) (001)-oriented (cubic notation) substrates subjected to different environmental treatments. We used total electron yield (TEY) spectroscopy probing a 4-5nm thin surface region [17] as well as fluorescence yield (FY) detection techniques known to be spectrally much more bulk sensitive. Spectra were measured at the Mn L-edge and at O K-edge. The former probe the unoccupied Mn 3d states via 2p ➔ 3d dipole transitions. The latter basically probe the unoccupied O 2p states via O 1s ➔ 2p transitions and indirectly give information on the Mn 3d occupancy and hence on the Mn valence due to the hybridization between the O 2p and Mn 3d orbitals [18]. In order to check whether the UHV or the synchrotron radiation was causing the Mn$^{2+}$ formation spectra were repeated 10 days later, the samples being kept in UHV during all this time.

## II. Experiment

About two months before the XAS synchrotron radiation experiments one set (named *A*) of three LCMO films was simultaneously grown (in order to have reproducible stoichiometrical, structural and physical properties) on LAO substrates by means of rf magnetron sputtering. During deposition the substrate temperature was kept at 800 °C. The pressure of the sputter gas was 330 mTorr (Ar- 20% O$_2$). Subsequently, films were in-situ annealed at 800 °C at an oxygen pressure of 350 Torr for 1 hour. Afterwards, they were cooled down to room temperature at a rate of 15°C/min at the same oxygen partial pressure. After finishing the growth process the three samples were kept under different environmental conditions. Sample *A$_{air}$* was kept in clean ambient atmosphere. Sample *A$_{vac}$* was held under vacuum conditions in a dry box (desiccator) to minimize air exposure; a residual exposure to ambient atmosphere of two days must be considered due to handling of the film. The third one (*A$_{anneal}$*) was annealed in air for two hours at T=1000ºC, with heating and cooling ramps of 5ºC/min, and thereafter also kept in air for about two months.

In order to study the time dependence of the spectral features a similar second set *B* of LCMO samples (*B$_{air}$*, *B$_{vac}$* and *B$_{anneal}$*) was prepared ten days before the synchrotron experiments.

A piece of the LCMO bulk target used for the growth of the films, with proper magnetic and structural properties of the La$_{2/3}$Ca$_{1/3}$MnO$_3$ composition, was prepared as further sample and measured as well its XAS spectrum to be used as reference and representative for Mn$^{3+}$:Mn$^{4+}$=2/3:1/3 valence ratio.

The thickness of the thin film samples and their surface roughness were deduced from grazing incidence x-ray reflectometry (XRR). The out-of-plane cell parameter $c$ was determined from X-ray diffraction (XRD) experiments using the (004) reflection. Magnetotransport properties were measured for films of set $A$. We used a four-probe configuration with a Quantum Design physical properties measurement system (PPMS) in the temperature range 10–300 K and with a maximum field of H=30 kOe applied perpendicular to the plane of the samples. Contacts were made by attaching platinum wires to the samples with silver paste. Magnetization curves, M(H) at T=10 K and M(T) with an applied field of H=5000 Oe, were measured by using a superconducting quantum interference device magnetometer (Quantum Design).

The XAS experiments were performed at the undulator beamline UE56-1-PGM-1 of the synchrotron radiation source BESSY [19]. The spectral resolution at the Mn 2p and O 1s edge was ca. E/ΔE=5000 and the degree of polarization was set to circular (right helicity and $P_{circ}$=0.85±0.03). We used the BESSY ultra-high vacuum polarimeter chamber [20], which allows the simultaneous measurement of TEY, FY and reflectivity. For the TEY detection the photo excited drainage current of the sample was recorded while the sample was kept at a potential of -95V with respect to the chamber. The fluorescence detector, a GaAsP photodiode was placed aside as close as possible to the sample. Nearby drainage electrodes were kept at a potential of +400 V. The angle of grazing incidence was fixed to $\phi_i = 40^o$, which is known from previous investigations to avoid saturation effects for the TEY data [21].

### III. Results
### IIIa. Structural and magnetic properties

The structural and magnetic properties of the samples $A_{air}$ and $A_{anneal}$ were measured ten days after the growth process (which includes the annealing at 1000°C for $A_{anneal}$ film). Sample $A_{vacuum}$ was measured after finishing the synchrotron experiments to minimize air exposure before the XAS measurements. Samples corresponding to set $B$ were also characterized after the XAS experiments. The Magnetotranport data were obtained after the synchrotron experiments in order to avoid surface contamination of the samples because of the silver paste used for the contacts.

A thickness $t$ of (21±1) nm and a *rms* surface roughness of (4±1) Å are obtained by fitting the XRR curves for the samples of set $A$. The thickness of set $B$ is $t$=(10±3) nm with indications of larger surface *rms* roughness. Samples $A_{air}$/$A_{vac}$ and $B_{air}$/$B_{vac}$ have larger out-of-plane cell parameter $c$=(3,945±0.005) Å and $c$=(3,950±0.005) Å, respectively, as the LCMO bulk (3.860 Å) which is considered to be due to the compressive strain induced by the in-plane mismatch with the substrate (3.79 Å). This is in agreement with previous results [22]. The annealed samples known to release part of their structural strain by an increase of the in-plane cell parameter relax to $c=(3.883\pm0.002)$ Å and $c=(3.930\pm0.002)$ Å for $A_{anneal}$ and $B_{anneal}$, respectively, approaching the bulk value.

The magnetotransport data indicate a large granularity for as grown films, but not for the annealed ones; the low temperature (T=10K) magnetoresistance ($MR_{10K}$) defined as normalized difference $(\rho_{H=0T}-\rho_{H=3T})/(\rho_{H=0T}+\rho_{H=3T})$, of the resistivity at zero field and high magnetic field, amounts to 37% for the $A_{vac}$ film and 46% for the $A_{air}$ one, while reaches only 1% for the $A_{anneal}$. In addition, the residual resistivity $\rho_{H=0T}$ at T=10K for $A_{vac}$ and $A_{air}$ is larger by a factor 10 and 20, respectively than that measured for $A_{anneal}$. The granularity of the as grown thin LCMO films is not unexpected because of the in-plane stress and the crystal properties of the LAO substrate [23]. For high temperature annealed films however the relaxation of the in-plane cell parameters improves the crystal quality of the LCMO layer.

The magnetization data for the two non-annealed samples of set $A$ (see inset Fig. 1) show that they exhibit a similar $T_c$ (≈223 K), clearly smaller than the bulk value (≈270 K). Nevertheless, large differences are found, when comparing their saturation magnetization $M_s$. While the $A_{vac}$ sample has a $M_s$ value of 570 emu/cm$^3$ only slightly different from that of bulk material (580 emu/cm$^3$), the $A_{air}$ film shows a significant reduction of it, $M_s$=503 emu/cm$^3$. This reduction of $M_s$ is even larger for the corresponding samples of set $B$ due to their smaller thickness [9]. The annealed samples of both sets, on the other hand, exhibit an epitaxial-like structure and a substantial improvement of their magnetic properties approaching those of the bulk material; for instance, sample $A_{anneal}$ exhibits $M_s$≈580 emu/cm$^3$ and $T_c$ ≈270 K. It is to be noted that sample $B_{anneal}$ still shows slightly depressed value of $T_C$ and $M_S$ which is correlated with a higher in-plane stress as inferred from the smaller enhancement of the $c$ cell parameter.

The discrepancies in the values of $MR_{10K}$, $\rho_{H=0T}$ and $M_s$ between the simultaneously grown films $A_{air}$ and $A_{vac}$ could be surprising but, as it will be shown later, are related to the period of time, during which $A_{air}$ sample was exposed to ambient atmospheric conditions before characterization of magnetic (10 days) and transport (2 months) properties. A complete report of the structural, magnetic and transport properties of these samples will be given elsewhere [24].

**IIIb. XAS**

Figure 2 and 3 show the TEY Mn L-edge spectra in the region of the $L_3$ and $L_2$ transitions, i.e. $2p_{3/2}$ and $2p_{1/2}$ hole states, for the 3 samples of set *A* and *B*, respectively. The spectra have been normalized at 643.2 eV where the bulk spectrum (also plotted for comparison) has its maximum intensity. We want to point out, however, that our essential conclusions are independent of the details of the normalization.

The spectra of the annealed samples $A_{anneal}$ and $B_{anneal}$ agree with each other and are essentially identical to that of the LCMO bulk sample. This means that the different period of air exposure, even on the time scale of several weeks, is unimportant for the annealed films. Thus, we conclude that they have achieved stable $Mn^{3+}$:$Mn^{4+}$ valence composition according to 2/3:1/3 ratio. In contrast, clear differences appear for the as grown films. Those kept in the desiccator ($A_{vac}$ and $B_{vac}$) also present almost identical spectra, i.e. are time independent. But careful inspection shows that they exhibit very slight differences at the low energy side of the $L_3$ and $L_2$ peaks, when compared to the bulk and annealed spectra. These differences have dramatically grown for the samples kept in air. While the spectrum corresponding to the $B_{air}$ film exposed for 10 days to ambient atmosphere presents a clear intensity increase at the low energy side of L-edges spectral features, a well defined sharp peak can be observed at 641.0 eV for the $A_{air}$ sample, kept in air for two months. This peak is comparable in size to the main peak at 643.2 eV and appears at the same position where only a shoulder can be inferred for the spectra of the $A_{vac}$ and $B_{vac}$ samples.

The fact that the spectrum of sample $A_{anneal}$ ($A_{vac}$) looks very much alike to that of sample $B_{anneal}$ ($B_{vac}$) also proves the reproducibility of the followed experimental procedure.

The insets of fig. 2 and 3 show various difference spectra, namely $A_{air}$ - $A_{anneal}$, $A_{vac}$ - $A_{ann}$ and $B_{air}$ - $B_{annead}$, $B_{vac}$ - $B_{anneal}$, in order to emphasize the observed changes in the spectra and to reveal their origin. For comparison a calculated divalent Mn ($3d^5$) spectrum [25] for a $Mn^{2+}$ ion in tetrahedral symmetry with a crystal field splitting of 0.5 eV is also shown. Excellent agreement of the relative peak positions, as well as of the spectral shape is found between this theoretical spectrum and the $A_{air}$ - $A_{anneal}$ difference. Moreover, other reported experimental $Mn^{2+}$ spectra [26-28] resemble our curve. Therefore, we claim that the extra features observed for the sample exposed to ambient atmosphere for 2 months are solely due to the presence of $Mn^{2+}$. For the normalized integrated divalent intensity contribution we obtain ($A_{air}$-$A_{anneal}$)/ $A_{anneal}$ = (16±1)%, i.e. we find that the Mn L-edge spectrum of the $A_{air}$ film can be decomposed into 84% due to stoichiometrically expected $Mn^{3+}$:$Mn^{4+}$=2/3:1/3 ratio and 16% due to $Mn^{2+}$. The same trend is observed for the $B_{air}$ sample. But here the $Mn^{2+}$ contributes only with (7±1)% to the total intensity (see inset of figure 3). The strong similarity between the $Mn^{2+}$ spectral contribution of $A_{air}$ and $B_{air}$ samples with the divalent Mn spectrum clearly indicates that the $3d^5$ electrons of such ions are localized an is at the origin of the large resistivity measured for the as grown films.

For the as grown samples kept under vacuum both the ($A_{vac}$ - $A_{anneal}$) and ($B_{vac}$ - $B_{anneal}$) difference spectra are identical in size and represent about (5±1)% of their respective integrated $A_{vac}$ and $B_{vac}$ spectra. The origin of such contribution is very likely due to the residual exposure to air. When normalizing all difference spectra to the maximum $L_2$ edge intensity (not shown), all curves resemble the divalent Mn spectrum except for a decrease of the peak at 640.9 eV for ($A_{vac}$ - $A_{anneal}$) and ($B_{vac}$ - $B_{anneal}$).

We present further support to the presence of $Mn^{2+}$ in as grown films kept in air. By the data in Fig. 4 obtained at the O K edge. The spectrum shows three main features at 544 eV, 536 eV and 530 eV, respectively. The origin of the broad peak at 544eV is attributed to electronic bands of Mn 4sp and La 6sp character, while the one at 536eV is related to bands of La 5d character [18]. The relevant peak for the study of a Mn valence changes is the one at 530 eV, including the high energy shoulder at 532.5 eV, which is due to dipole transitions from O 1s to O 2p states, which are hybridized with the unoccupied Mn 3d orbitals. The intensity of this peak represents the 2p hole and is also an indirect measure of the Mn 3d level occupancy. Fig. 4 shows the measured spectra for the 3 samples of set *A* as well as for the LCMO bulk reference sample. They are normalized to the same area assuming similar numbers of free states and similar oscillator strength. Similar results (not shown) are obtained for the three samples of set *B*. The size of the low energy peak of sample $A_{anneal}$ ($B_{anneal}$) is similar to that measured for the bulk sample indicating a similar occupancy and thus a similar $Mn^{3+}$:$Mn^{4+}$ valence ratio, consistent with the Mn L-edge data. A decrease of this peak is observed for the $A_{air}$ ($B_{air}$) film which reveals an increase of the occupancy of the Mn 3d levels [29], i.e. a decrease of the Mn valence with respect to the annealed and bulk values.

In order to determine whether the observed $Mn^{2+}$ spectral contribution is restricted to the outermost layers of the films as expected by the air exposure, we have compared the TEY spectra with those obtained by means of fluorescence yield detection. But our FY data show similar results at both edges (Mn and O) revealing that those "anomalies" are not exclusively confined close to the surface layers. In Fig. 5 we show a comparison between the TEY and FY data for the $A_{air}$ sample at the Mn L-edge. Differences with respect to the $L_3/L_2$ ratio can be noticed after normalization, most likely due to saturation effects [30] for the FY data at the selected angle of incidence.

The repetition of all spectra 10 days later, without breaking the UHV, showed identical features as those already commented on thus, ruling out the UHV and the synchrotron radiation as a factors responsible for the $Mn^{2+}$ formation. The differences of thickness and/or magnetic properties between the two sets of samples can also be ruled out. Similar XAS experiments on a LCMO film with lower thickness (6nm) and lower magnetic properties ($T_c$ =160 K and $M_S$=110 emu/cm$^3$) than these measured for $B_{air}$ showed a $Mn^{2+}$ integrated spectral contribution of (16±1)% after 1 month of air exposure. Similar to that found for the $A_{air}$ film and larger than that of the $B_{air}$ one. Therefore the spectral differences found between the films can only be related to their different exposure time to air.

From the comparison of data corresponding to samples of series *A* and *B* the temporal evolution of the Mn valence balance for an as grown LCMO/LAO thin film exposed to ambient air can be deduced (figure 6). Unfortunately we cannot perform *in situ* XAS measurements which preclude determination of the actual Mn valence of as grown samples prior to venting the evaporation chamber. The similarity between the spectra of $A_{vac}$ and $B_{vac}$ films in spite of their different age indicates that the vacuum environment of the desiccator stabilizes the Mn valence avoiding the degradation of the films. This fact, together with the consideration that due to the handling of the samples they were exposed about 2 days to ambient atmosphere indicates that within this period of time the degradation of the film has already started, a 5% of the total Mn L-edge intensity. After 10 days of air exposure ($B_{air}$) the degradation continues and a clear $Mn^{2+}$ component representing a 7% of the total intensity is present. The $Mn^{2+}$ formation goes on and two months later such a component has increased up to 16% (see fig. 6).

Annealing of the as grown samples in air at 1000°C for two hours recovers bulk-like properties (cell parameter, $M_s$, $T_c$) and the nominal $Mn^{3+}$:$Mn^{4+}$=2/3:1/3 ratio with no $Mn^{2+}$ presence. The Mn valence of these films is stable towards air exposure at room temperature, at least for a two months period.

It is worth mentioning at this point that differences in $M_s$ between $A_{vacuum}$ and $A_{air}$ were already observed prior to the XAS measurements. Since the films were grown simultaneously it would be reasonable to expect the same magnetic behaviour for both. Nevertheless, as mentioned previously, the magnetic measurements before the XAS experiments for the $A_{air}$ sample were done 10 days after its growth, thus after ten days of air exposure. Since the spectrum of the $B_{air}$ sample clearly demonstrates that within this period of time the $Mn^{2+}$ formation has already started, we think that this non expected divalent Mn component is the responsible of such differences. In order to confirm it we repeated structural and magnetic characterization for the *A$_{air}$* film after the XAS experiments, i.e. after two months of air exposure. Neither changes of the out-of plane cell parameter nor in the transition temperature Tc were observed but, as shown in Fig. 1 (red line), a further reduction of M$_s$ was found concomitant with the increase of the Mn$^{2+}$ contribution pointing to a non-ferromagnetic order of such component. The formation of divalent Mn and the localization of its five 3d electrons also explains the differences in MR$_{10K}$ and ρ$_{H=0T}$ found between *A$_{vacuum}$* and *A$_{air}$* since as commented the transport data was obtained after two months of air exposure for the *A$_{air}$* film, i.e. presenting 16% of Mn$^{2+}$ integrated spectral contribution.

### IV. Discussion

We note that similar spectral features in mixed valence manganites, namely an increase in intensity at the low energy side of the Mn L-edges, have been reported earlier, both theoretically [25, 31] and experimentally [32]. They were attributed to strain induced change in the crystal field strength and/or by a change in the Mn$^{3+}$:Mn$^{4+}$ ratio without involving another valence state of Mn. These earlier findings together with the probable believe in the stability of the trivalent and tetravalent Mn valences in these materials have previously prevented to look for other plausible explanations. In our case, the observed differences between the *A$_{air}$* and *B$_{air}$* as opposed to the *A$_{vac}$* and *B$_{vac}$* ones strongly suggest that other possibilities have to be considered. Our results shows that a degradation of LCMO thin films takes place in samples exposed to ambient atmosphere and that this degradation increases with time exposure. Interestingly we also find that a stabilization of the Mn valence balance is accomplished by high temperature (1000°C) annealing in air. We also find that samples kept in vacuum are stabile whereas exposure to air contributes to Mn$^{2+}$ formation.

In order to explain the Mn$^{2+}$ formation Hundley et al [15] and later on de Jong et al. [16] suggested an instability of Mn$^{3+}$ to be responsible of the Mn$^{2+}$ formation via 2Mn$^{3+}$→Mn$^{2+}$ + Mn$^{4+}$. Such a mechanism cannot be invoked in the present case, since the Mn L-edge spectra of the as grown and air exposed *A$_{air}$* and *B$_{air}$* samples are completely explained with the solely combination of two spectral components; one being due to a correct Mn$^{3+}$:Mn$^{4+}$=2/3:1/3 stoichiometric ratio and the other one corresponding to a Mn$^{2+}$ contribution. Thus, no excess of Mn$^{4+}$, as would be expected from the above reaction, is detected.

Since the Mn$^{2+}$ component develops and increases due to air exposure a reducing media present in air must be at the origin of the Mn reduction by contributing electrons to the LCMO system and/or by removing oxygen. To this respect R. Cracium et al. have demonstrated that in $MnO_x$-YSZ catalytic materials, $Mn^{n+}$ can provide sites for CO adsorption and supply oxygen from its oxide structure for oxidation, effectively leading to $Mn^{2+}$ [33]. A similar mechanism might be at the origin of the $Mn^{2+}$ formation in our films.

An increase of the free surface in direct contact with air should produce an increase of the $Mn^{2+}$ presence. Thus its formation should be strongly enhanced in films with a marked granular character, this appears to be the case for non-annealed samples of the present study. Granularity also explains the presence of $Mn^{2+}$ not only for regions close to the surface but also within the films. This is in agreement with similar measurements [34] from a set of LCMO films grown on top of $SrTiO_3$ and $NdGaO_3$ (NGO) substrates (with 1% and almost zero lattice mismatch, respectively). A clear $Mn^{2+}$ signature was observed for films grown on NGO, known by atomic force microscopy images to present a strong granularity. On the other hand, in high quality epitaxial thin films as those grown on STO substrates [9] the effect is likely to be restricted to an area close to the surface [16]. This has in fact been corroborated by Hall effect measurements in as grown and annealed LCMO/STO epitaxial thin films [35].

Interesting enough, the existence of a region close to the free surface with presence of $Mn^{2+}$ with localized $3d^5$ electrons can explain why surface resistance is larger in as grown than in annealed films, as observed in AFM current sensing measurements in LCMO/LAO samples [36]. Our results also offer some clues to better understand effects such as the reduction of spin polarization close to the film surface and its faster decrease with temperature [37].

The $Mn^{2+}$ ion formation is not exclusively detected in $La_{2/3}Ca_{1/3}MnO_3$ thin films. Similar observations were also made in $La_{0.5}Ca_{0.5}MnO_3$ samples [38]. Furthermore, as commented by de Jong et al. the presence of $Mn^{2+}$ might have previously been incorrectly related to an increase of $Mn^{4+}$ as well as to a change of the crystal field strength in both $La_{1-x}Ca_xMnO_3$ and $La_{1-x}Sr_xMnO_3$ films. We believe that formation of divalent Mn can be a general feature occurring on Lanthanum-Manganese perovskites, particularly when exposed to ambient atmosphere.

## V. Conclusions

$La_{2/3}Ca_{1/3}MnO_3$ thin films grown on $LaAlO_3$ substrates and subjected to a high temperature (1000°C) annealing process (in air), present bulk-like spectral properties even if they were subsequently exposed for prolonged period of time to ambient atmosphere. Non-annealed samples exposed to ambient atmosphere give rise to the appearance of $Mn^{2+}$ in addition to the expected trivalent and tetravalent Mn components. Concomitant with this, such samples exhibit reduced magnetic and magnetotransport properties, as expected from the localization of the five 3d electrons, in comparison with bulk material. This reduction and the $Mn^{2+}$ abundance increase after a long lasting air exposure. In contrast, samples kept under vacuum conditions (dry box) do not show this aging process, provided they are not exposed to air.

Similar Mn spectral features as those we report have been observed in $La_{1/2}Ca_{1/2}MnO_3$ as well in $La_{2/3}Sr_{1/3}MnO_3$ compounds, thus we believe that the formation of divalent Mn due to air exposure is a common feature of manganite oxides. The wide-spread believe in the Mn valence stability of the mixed-valence compounds may be responsible that any reduction of the magnetic and/or transport properties of manganite-based CMR type devices is explained in physical terms where only $Mn^{3+}$ and $Mn^{4+}$ ions are invoked. Similar considerations were already put forward to interpret XAS spectra. We believe that our results are of major relevance and point out the necessity of also taking into account the possible presence of $Mn^{2+}$ in order to explain other previous experimental data, especially when the films have been exposed to ambient atmosphere. Further investigations are, however, needed to clarify the Mn valence balance control in free surfaces and interfaces. Since these materials do play an important role on the implementation of oxide-based thin films magnetoresistive devices such investigations are not only of pure scientific interest.

**Acknowledgments**

One of the authors (S. Valencia) thanks Dr. L. Soriano for valuable discussions. Financial support from the MCyT (Spain) and FEDER (EC) project MAT2003-04161 is acknowledged.

**Figure captions**

Figure 1. (Color online) Magnetic hysteresis cycles (right) measured at T=10K. The results for the $A_{air}$ sample measured after the XAS experiments. i.e. two months after the growth, are shown with a continuous line. Inset: Magnetization curve as function of temperature measured with H=5000 Oe

Figure 2. (Color online) Manganese L-edge XAS spectra measured by TEY for the samples $A_{air}$, $A_{vac}$ and $A_{anneal}$ grown 2 months before experiments. The spectrum for a bulk reference sample (red continuous line) is shown for comparison. The inset shows the differences ($A_{air}$ - $A_{anneal}$) and ($A_{vac}$ - $A_{anneal}$). The theoretical spectrum for $Mn^{2+}$ is also plotted (continuous line). The presence of $Mn^{2+}$ is evident in the as grown sample expose to air

Figure 3. (Color online) Manganese L-edge XAS spectra measured by TEY for the samples $B_{air}$, $B_{vac}$ and $B_{anneal}$ grown 10 days before experiments. The spectrum for a bulk reference sample (red continuous line) is shown for comparison. The inset shows the differences ($B_{air}$ - $B_{anneal}$) and ($B_{vac}$ - $B_{anneal}$). For comparison the theoretical spectra for $Mn^{2+}$ is also plotted. Some $Mn^{2+}$ has occurred after exposure to air.

Figure 4. (Color online) Oxygen K-edge XAS spectra measured by TEY for the samples $A_{air}$, $A_{vac}$ and $A_{anneal}$, normalized to the area. As a reference the LCMO bulk spectrum measured by FY is also shown (dash line).

Figure 5. Comparison between the Mn L-edge spectra obtained by means of TEY (filled dots) and FY (open dots) for the sample kept in air $A_{air}$.

Figure 6. Main panel: Temporal evolution of the divalent Mn component. Inset: Temporal evolution of the saturation magnetization for the $A_{air}$ sample.

**Figure 1**

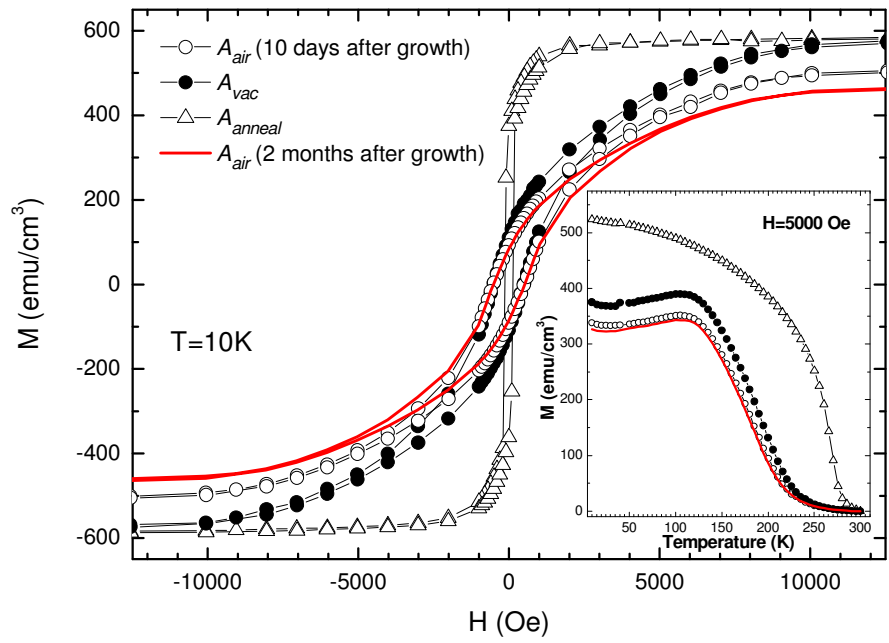

**Figure 2**

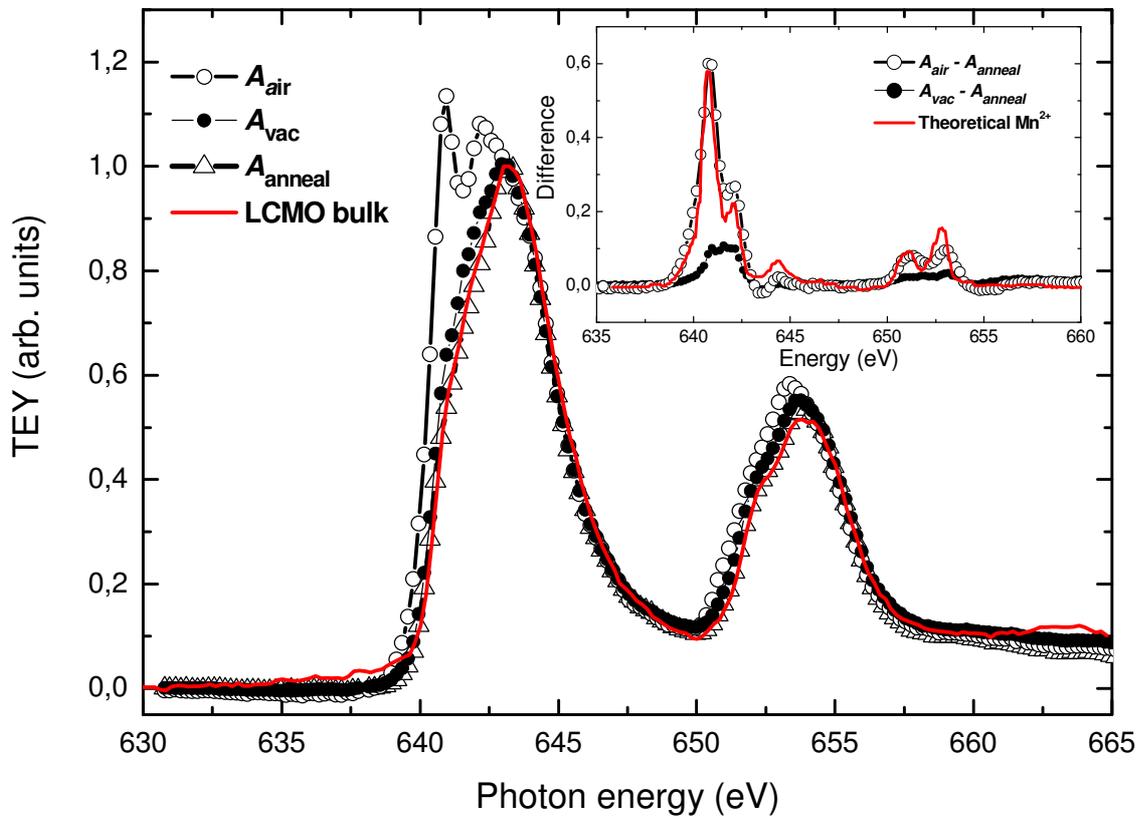

**Figure 3**

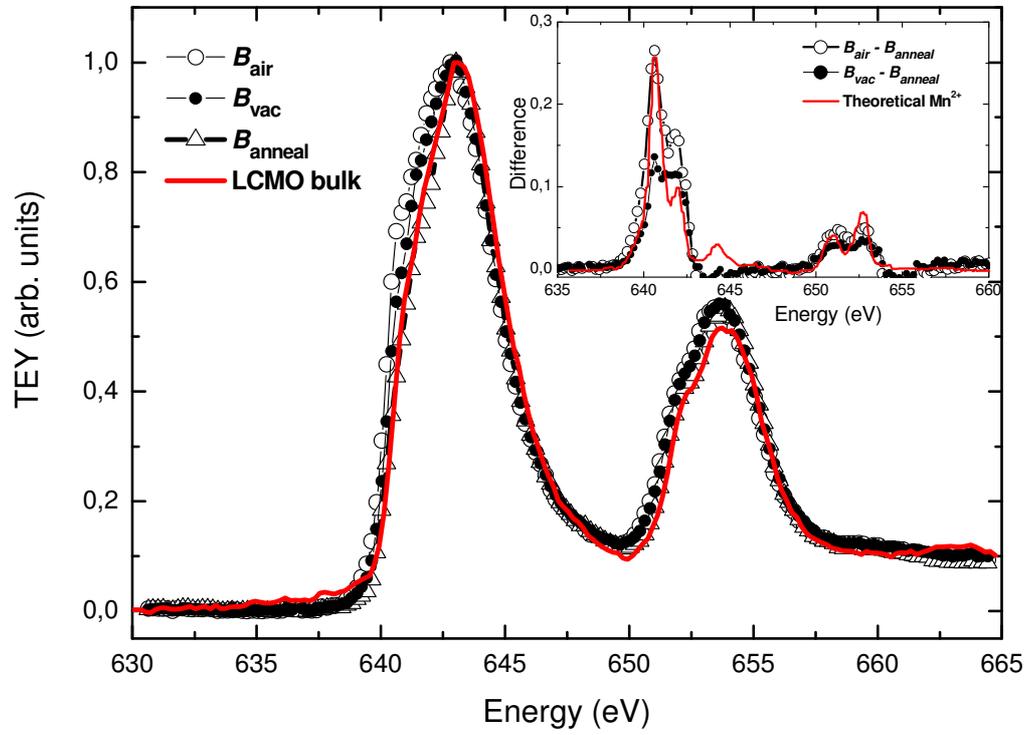

**Figure 4**

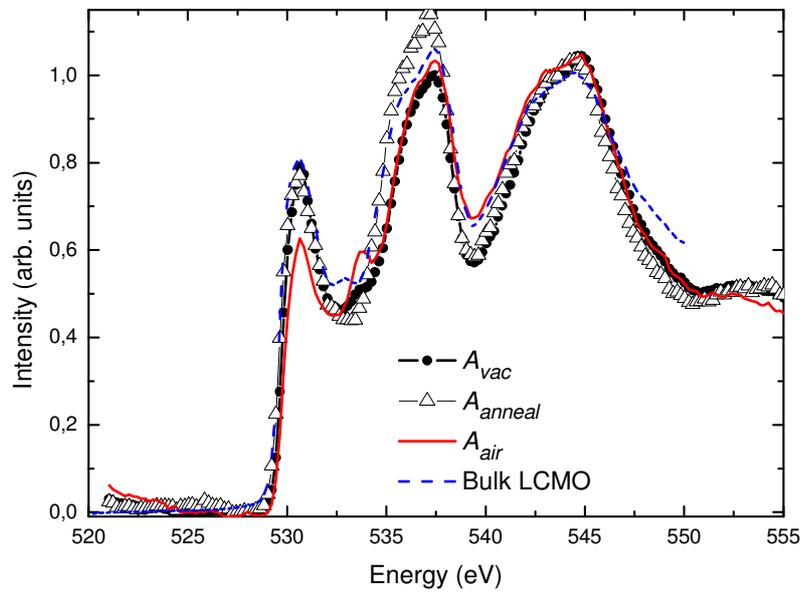

**Figure 5**

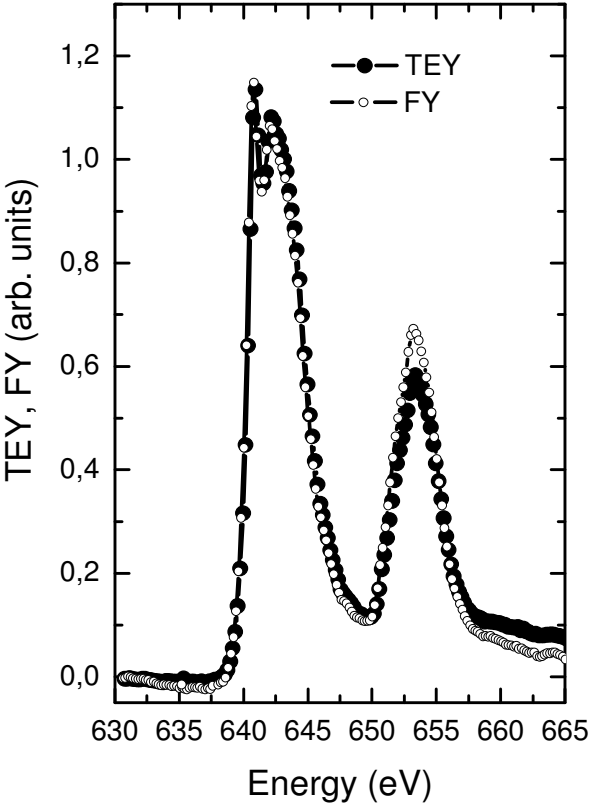

**Figure 6**

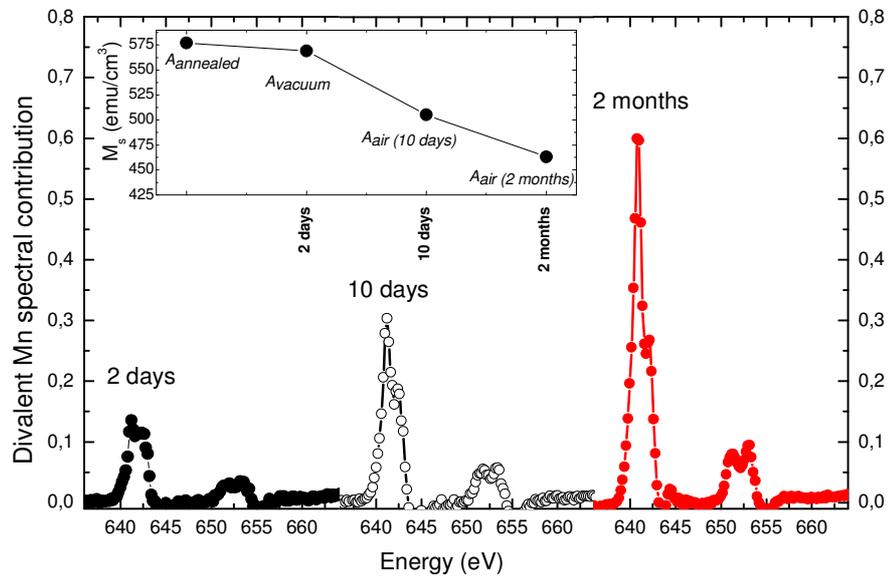